\documentclass{PoS}

\title{Westerbork ultra-deep survey of HI at z=0.2}

\ShortTitle{Westerbork survey of HI emission at z=0.2}

\author{\speaker{Marc Verheijen}$^{,a}$, Boris Deshev$^a$,\hspace{10cm} Jacqueline van Gorkom$^b$, Bianca Poggianti$^c$, Aeree Chung$^d$, Ryan Cybulski$^e$,\hspace{1cm} K.S. Dwarakanath$^f$, Mar\'ia Montero-Casta\~no$^b$, Glenn Morrison$^g$,\hspace{3cm} David Schiminovich$^b$, Arpad Szomoru$^h$, Min Yun$^e$ \\
\llap{$^a$} Kapteyn Astronomical Institute, University of Groningen, Groningen, The Netherlands\\
\llap{$^b$} Department of Astronomy, Columbia University, New York, U.S.A.  \\
\llap{$^c$} Istituto Nazionale di Astrofisica (INAF), Osservatorio Astronomico di Padova, Padova, Italy \\
\llap{$^d$} Harvard-Smithsonian Center for Astrophysics, Cambridge, Massachusetts, U.S.A. \\
\llap{$^e$} Department of Astronomy, University of Massachusetts, Amherst, Massachusetts, U.S.A. \\
\llap{$^f$} Raman Research Institute, Bangalore, India \\
\llap{$^g$} Institute for Astronomy, University of Hawaii, Honolulu, Hawaii, U.S.A. \\
\llap{$^h$} Joint Institute for VLBI in Europe (JIVE), Dwingeloo, The Netherlands \\
E-mail: \email{verheyen@astro.rug.nl}}

\abstract{In this contribution, we present some preliminary
observational results from the completed ultra-deep survey of 21cm
emission from neutral hydrogen at redshifts z=0.164$-$0.224 with the
Westerbork Synthesis Radio Telescope. In two separate fields, a total
of 160 individual galaxies has been detected in neutral hydrogen, with
HI masses varying from 1.1$\times$10$^{9}$ to 4.0$\times$10$^{10}$
M$_{\odot}$.  The largest galaxies are spatially resolved by the
synthesized beam of 23$\times$37 arcsec$^2$ while the velocity
resolution of 19 km/s allowed the HI emission lines to be well
resolved. The large scale structure in the surveyed volume is traced
well in HI, apart from the highest density regions like the cores of
galaxy clusters. All significant HI detections have obvious or
plausible optical counterparts which are usually blue late-type
galaxies that are UV-bright. One of the observed fields contains a
massive Butcher-Oemler cluster but none of the associated blue
galaxies has been detected in HI. The data suggest that the
lower-luminosity galaxies at z$\approx$0.2 are more gas-rich than
galaxies of similar luminosities at z=0, pending a careful analysis of
the completeness near the detection limit. Optical counterparts of the
HI detected galaxies are mostly located in the 'blue cloud' of the
galaxy population although several galaxies on the 'red sequence' are
also detected in HI. These results hold great promise for future deep
21cm surveys of neutral hydrogen with MeerKAT, APERTIF, ASKAP, and
ultimately the Square Kilometre Array. }

\FullConference{ISKAF2010 Science Meeting - ISKAF2010\\
		June 10-14, 2010\\
		Assen, the Netherlands}

\begin{document}

\section{Introduction}

\begin{figure}[t]
\includegraphics[width=\textwidth]{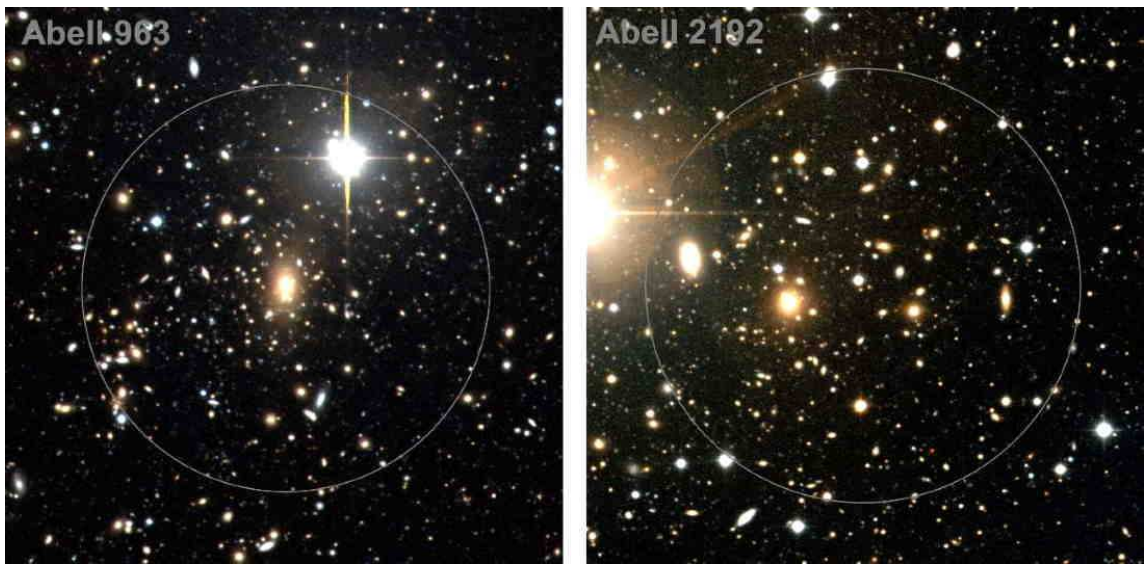}
\caption{Optical images of the galaxy clusters Abell 963 at z=0.206
(left) and Abell 2192 at z=0.188 (right). The colours are constructed
from deep B- and R-band images aquired with the Wide Field Camera of
the Isaac Newton Telescope. Although the entire 1 deg$^2$ of the
primary beam of the WSRT was imaged, these images only show an area of
6.8$\times$6.8 arcmin$^2$ centered on the clusters. The large circles
have a diameter of 1 Mpc at the distances of these clusters.}
\label{fig1}
\end{figure}

It is important to understand the role and fate of neutral hydrogen as
a basic ingredient of the formation process of galaxies and as a
sensitive tracer of subsequent evolutionary processes, both over
cosmic time and in different environments. In the local universe, the
morphology and kinematics of the cold atomic gas in and around
galaxies often reveal physical mechanisms such as tidal interactions,
ram-pressure stripping and 'cold accretion' that are very difficult to
detect by other observational means (e.g. van der Hulst, this
issue). Furthermore, these mechanisms are strongly influenced by the
particulars of the local and global environments in which galaxies
reside.

The predominance and efficiency of such processes are likely to evolve
over cosmic time. For instance, the overall star formation density in
the universe has declined by an order of magnitude in the last 6 Gyr
while the morphological mix of galaxy populations has changed
significantly, especially in the dense environments of galaxy
clusters. The fraction of blue galaxies in clusters was higher in the
past (e.g. Butcher \& Oemler, 1978) while the relative number of
spiral galaxies in clusters increases with redshift, at the expense of
the S0 population (e.g. Fasano et al 2000).

At a redshift of 0.2 or a look-back time of 2.5 Gyr, signatures of
cosmic evolution of the galaxy populations become evident. The nearest
Butcher-Oemler (B-O) clusters are found at z$\approx$0.2 and one of
the many questions one could ask, is whether the B-O effect relates to
the rate at which galaxy clusters accrete their members or whether it
is caused by the accreted population of field galaxies that may be
more gas-rich as these redshifts. To address this issue and many
others, like the cosmic evolution of the HI Mass Function and
$\Omega_{\rm HI}$, we set out to study the HI properties of galaxies
at z$\approx$0.2 with the Westerbork Synthesis Radio telescope (WSRT).
The goal is to relate the HI characteristics of the galaxies to their
luminosities, morphologies, star formation rates, the evolutionary
state of their constituent stellar populations, as well as to their
global and local environments.

\begin{figure}[t]
\includegraphics[width=\textwidth]{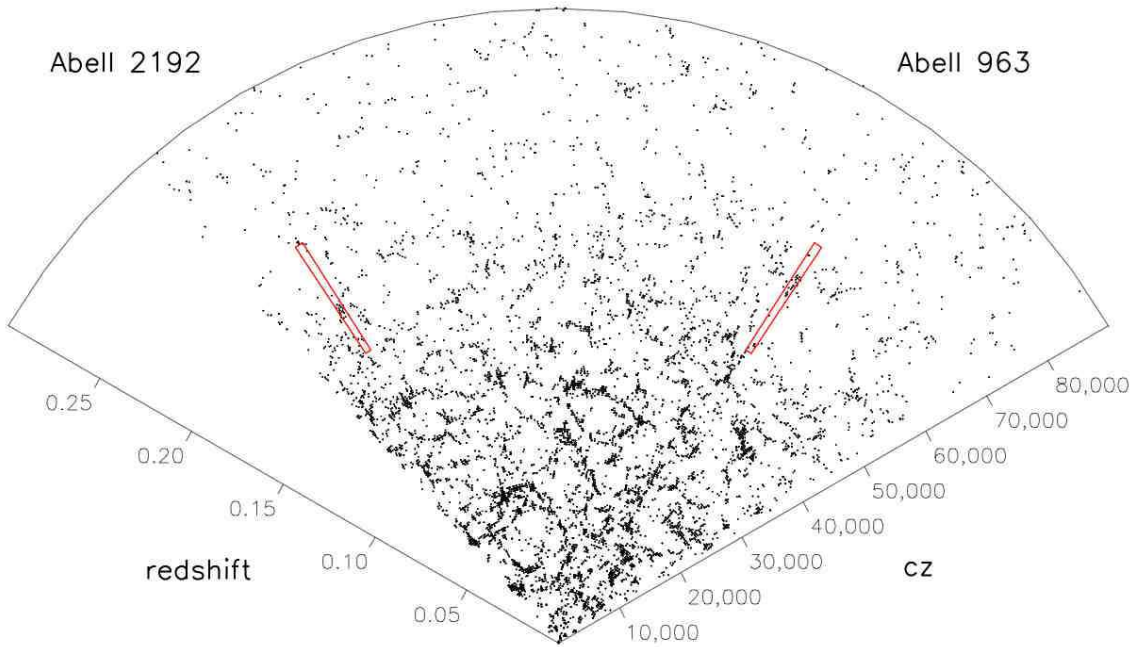}
\caption{A redshift pie-diagram along a great circle passing through
both clusters. Dots represent galaxies with optical redshifts from the
SDSS. The elongated rectangles indicate the two volumes that have been
surveyed with the WSRT, covering the redshift range
0.164$-$0.224. These volumes sample both overdense and underdense
regions. At these redshifts, the large scale sctructure becomes
sparsely sampled by the SDSS. Note that the HIPASS survey does not
extend beyond cz=12,000 km/s while the Alfalfa survey with the Arecibo
telescope is restricted to z$\lesssim$0.06.}
\label{fig2}
\end{figure}

\section{Target fields}

Two single-pointing fields were selected in the northern sky, each
containing an Abell cluster of galaxies and the large scale structure
in which they are embedded, as well as foreground and/or background
voids, thus sampling the broadest range of cosmic environments. Abell
963 at z=0.206 is a massive lensing cluster with a strong B-O effect
and an unusual large fraction of blue galaxies (f$_{\rm B}$=0.19,
Butcher et al 1983). With a velocity dispersion of 1350 km s$^{\rm
-1}$ it is a bright X-ray source and the regular X-ray contours,
centered on a central cD galaxy, suggest a low level of substructure
in this cluster. Abell 2192 at z=0.188 is less massive and more
diffuse. The fraction of blue galaxies is yet unknown for this
cluster. With a velocity dispersion of 650 km s$^{\rm -1}$, this
cluster is barely detected in X-rays. Figure 1 shows a color image of
both clusters. It should be stressed that the galaxy clusters
themselves only occupy $\sim$4\% of the surveyed volumes.

Figure 2 indicates the locations of the two volumes in a pie-diagram
of redshift space along a great circle through both
clusters. Redshifts are taken from the SLOAN survey. Clearly, based on
optical redshifts, the large scale structure is less well sampled at
these distances. Note that the blind HIPASS survey does not extend
beyond cz=12,000 km s$^{\rm -1}$ and that the Alfalfa survey with the
Arecibo telescope is limitd to a maximum redshift of
z$\approx$0.06. The widths and depths of the volumes surveyed with the
WSRT are explained in the next section.

\begin{figure}[t]
\includegraphics[width=\textwidth]{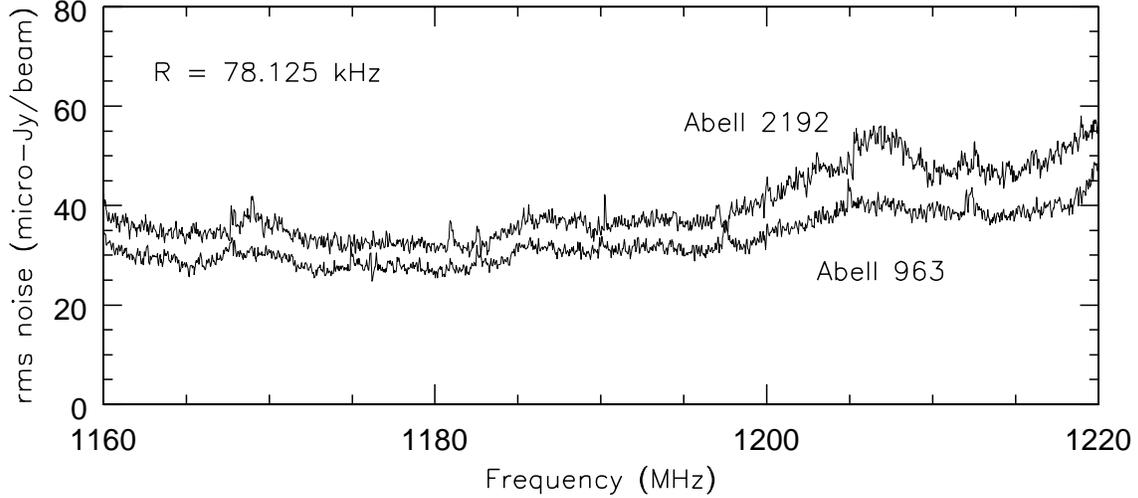}
\caption{The achieved rms noise levels in the continuum subtracted
data cubes as a function of frequency, after Hanning smoothing was
applied to yield a FWHM spectral resolution of 78.125 kHz. Although
the data for both fields was calibrated and processed independently,
the rms noise in both fields shows the same trends with
frequency. This is likely the result of low-level RFI that could not
be removed from the individual measurements.These noise levels are
close to the expected thermal noise.}
\label{fig3}
\end{figure}

\section{Observations}

The observations took maximum advantage of the powerful IF system and
backend of the WSRT. Eight overlapping IF bands, each 10 MHz wide and
with 256 dual-polarisation channels, covered the contiguous frequency
range between 1220 and 1160 MHz with 1236 channels, each 39.0625 kHz
wide. This corresponds to a redshift range of 0.1643$<$z$<$0.2245 or
49,426$<$cz$<$67,300 km s$^{\rm -1}$. For $\Omega_{\rm M}$=0.3,
$\Omega_{\Lambda}$=0.7 and H$_0$=70 km s$^{\rm -1}$Mpc$^{\rm -1}$,
this corresponds to a range in luminosity distance of 789$<$D$_{\rm
lum}$$<$1,117 Mpc or an effective volume depth of 328 Mpc. At 1190
MHz, the Full-Width-Quarter-Maximum of the primary beam of the WSRT is
61.7 arcminutes wide, or 11.9 Mpc. Consequently, the entire combined
volume that has been blindly surveyed within the 60 MHz passband and
the FWQM of the primary beam is $\sim$73,000 Mpc$^3$. This is
equivalent to the volume of the Local Universe within a distance of 26
Mpc. After identifying the HI emission and cleaning of the channel
maps, the synthesized beam was restored with a Gaussian beam with a
FWHM of 23$\times$37 arcsec$^2$ which corresponds to 74$\times$119
kpc$^2$ at the center of the band.

\begin{figure}[ht]
\includegraphics[width=0.94\textwidth]{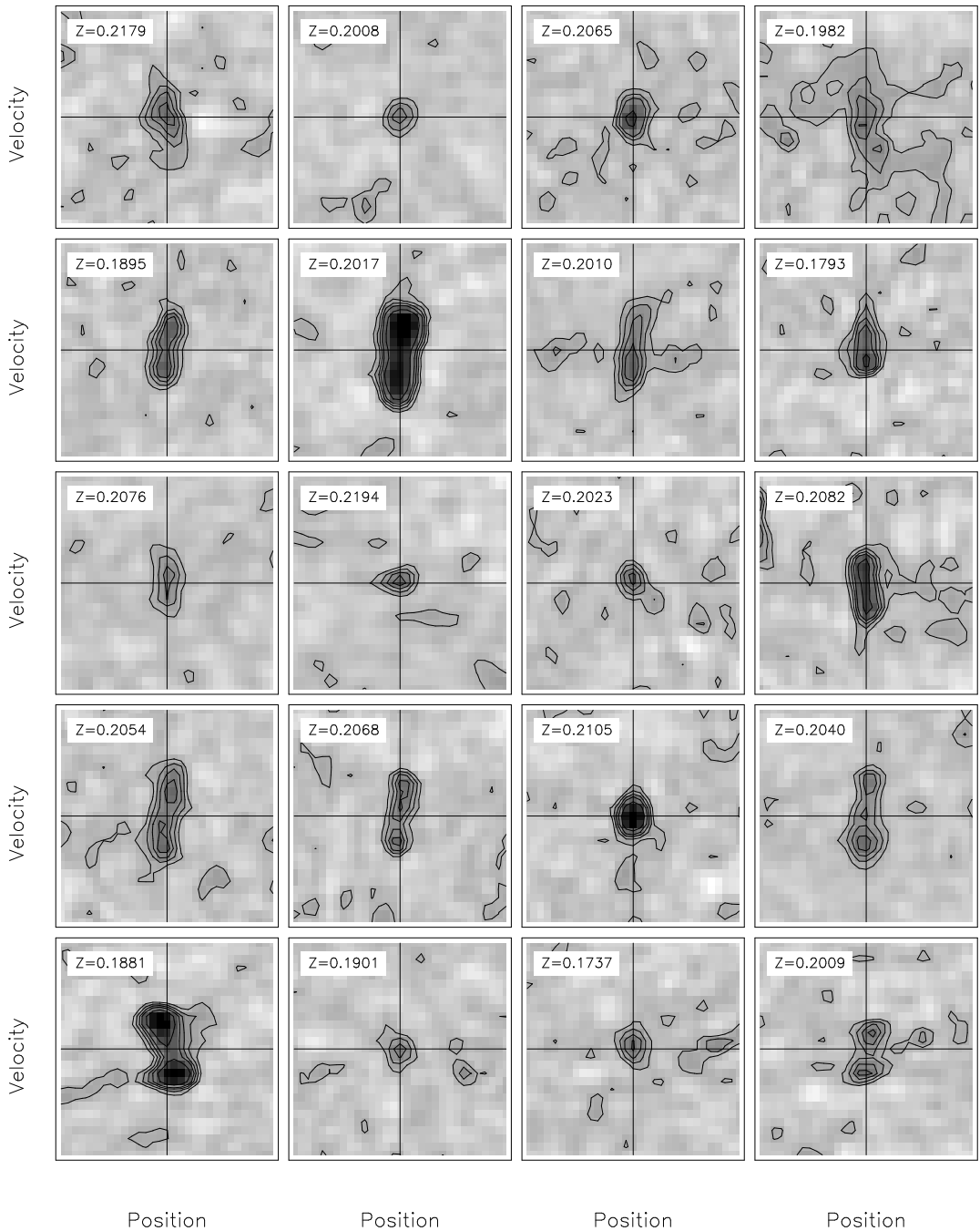}
\caption{Position-velocity diagrams of 20 random HI detections. The
vertical velocity axis covers $\sim$1,000 km/s in the rest-frame of
the galaxies. The horizontal axis covers 1 arcminute on either side of
the source. Horizontal lines indicate the systemic velocity, the
vertical lines correspond to the central position of the HI
source. Contour levels are at 1.5, 3.0, 4.5, 6.0, 9.0 and 12.0 times
the local noise. The FWHM spectral resolution in these maps is 312.5
kHz, or 76 km/s at the center of the band.  Not all sources fullfil
the detection criteria at this particular spectral resolution. Note
that many galaxies show the typical double-peaked velocity profiles
and that several galaxies are spatially resolved.}
\label{fig4}
\end{figure}

The goal of these ultra-deep HI observations is to be able to detect,
at the distance of each cluster, a minimum HI mass of 2$\times$10$^9$
M$_\odot$ over a 150 km s$^{\rm 1}$ wide emission line, with a
signal-to-noie of 4 in each of three adjacent and independent spectral
resolution elements. This corresponds to a limiting column density of
3$\times$10$^{19}$ cm$^{-2}$ over a 80 km s$^{\rm -1}$ wide profile at
the 7-sigma level. These detection limits require a total integration
time of 78$\times$12$^{\rm hr}$ for Abell 2192 and 117$\times$12$^{\rm
hr}$ for Abell 963. The observations were collected in a 8-semester
long-term Large Program during the years 2005 through 2008. First
results from the pilot observations in 2005 are described by Verheijen
et al. (2007). 

Figure 3 shows the achieved rms noise in each continuum-subtracted
data cube as a function of frequency, after Hanning smoothing to a
spectral resolution of 78.125 kHz. The variations in the rms noise are
a consequence of RFI residuals and frequency dependent flagging,
especially above 1200 MHz. Below 1200 MHz, 5-8\% of the visibilities
were flagged but above 1200 MHz this was 15-17\%. The measured noise
levels are very close to the thermal noise that could be expected
given the number of retained visibilities and the effective A/T$_{\rm
sys}$ of the telescope. At these extremely low noise levels,
imperfections in the bandpass calibration become noticable after
combining the individual measurements. In particular the
position-dependent shape of the bandpass results in significant
frequency-dependent residuals after subtraction of the continuum
sources. Fortunately, these residuals have sufficient spatial and
frequency coherence to avoid confusion with real sources, but they do
affect the overall rms noise in a channel map. The non-uniformity of
the noise, however, does complicate automatic source finding
algorithms and corrections for incompleteness.

The deep HI observations with the WSRT are supplemented with B- and
R-band imaging with the INT, NUV and FUV imaging with GALEX, 3.6, 4.5,
5.6, 8, 24 and 70 micron imaging with Spitzer, as well as optical
spectroscopy with WIYN. This contribution, however, focuses on the HI
observations and the optical luminosities and colours of the HI
detected galaxies.

\begin{figure}[t]
\includegraphics[width=\textwidth]{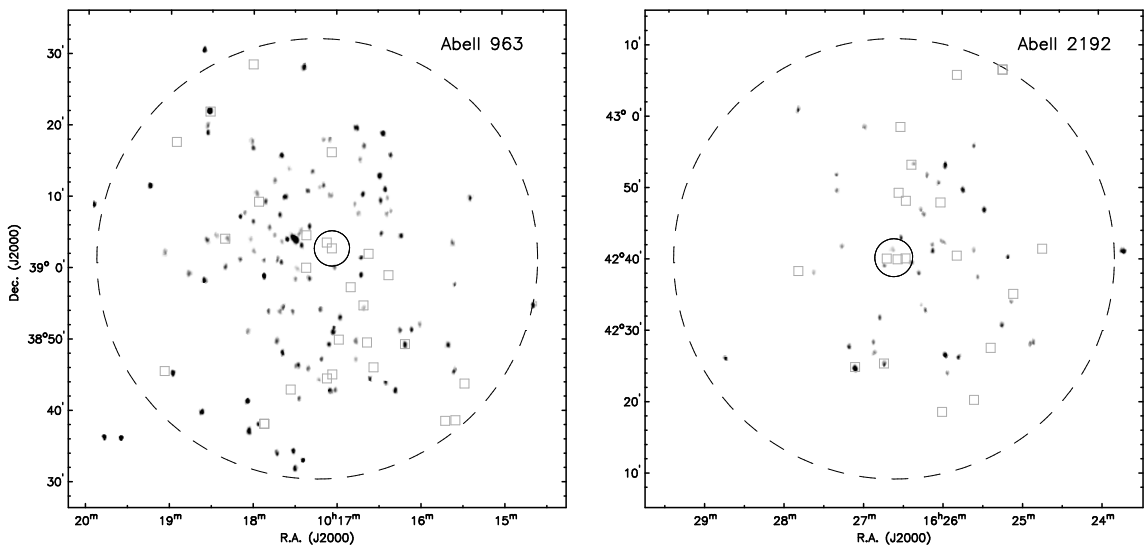}
\caption{HI column density maps of the entire WSRT field-of-view,
integrated over the full redshift range. HI detections are indicated
by grayscales. Open squares indicate the positions of galaxies with
optical redshifts from the SDSS that fall within the 0.164$-$0.224
redshift range. There is very little correspondence between the HI
detected galaxies and the galaxies targeted for optical spectroscopy
by the SDSS. The large dashed circle indicates the FWQM of the primary
beam of the WSRT. The small circles near the center correspond to the
circles in Figure 1.}
\label{fig5}
\end{figure}

\section{Preliminary results}

After initial Hanning smoothing, the data cubes were smoothed further
to a Gaussian frequency response with a FWHM of 4, 6 and 8 spectral
channels. Thus, four data cubes exist for each field, with FWHM
spectral resolutions of 78.125, 156.25, 234.375 and 312.5 kHz or 19,
38, 57 and 76 km/s at 1190 MHz.

Sources were automatically identified by searching in all data cubes
for pixels above a certain threshold that are connected in frequency
space. Given the structure in the noise, detection criteria were set
conservatively at 1$\times$8$\sigma$, 2$\times$5$\sigma$,
3$\times$4$\sigma$, or 4$\times$3$\sigma$. In the latter case, four
independent but connected spectral resolution elements, each at least
3$\sigma$ above the noise in a channel map, would constitute a
detection. Each detection was visually inspected to avoid confusion
with structure in the noise. In several cases, only one peak of a
double-horned profile was detected or only part of an asymmetric HI
profile was isolated by this detection algorithm. After removing
spurious detections, correcting incomplete detections and verifying
the presence of an optical and UV counterpart, a total of 160 galaxies
was detected in HI, 118 in Abell 963 and 42 in Abell 2192. The
significantly different detection rates for both fields can be
ascribed to a difference in the rms noise levels, combined with cosmic
variance of the large scale structure in the two volumes.

\begin{figure}[t]
\includegraphics[width=\textwidth]{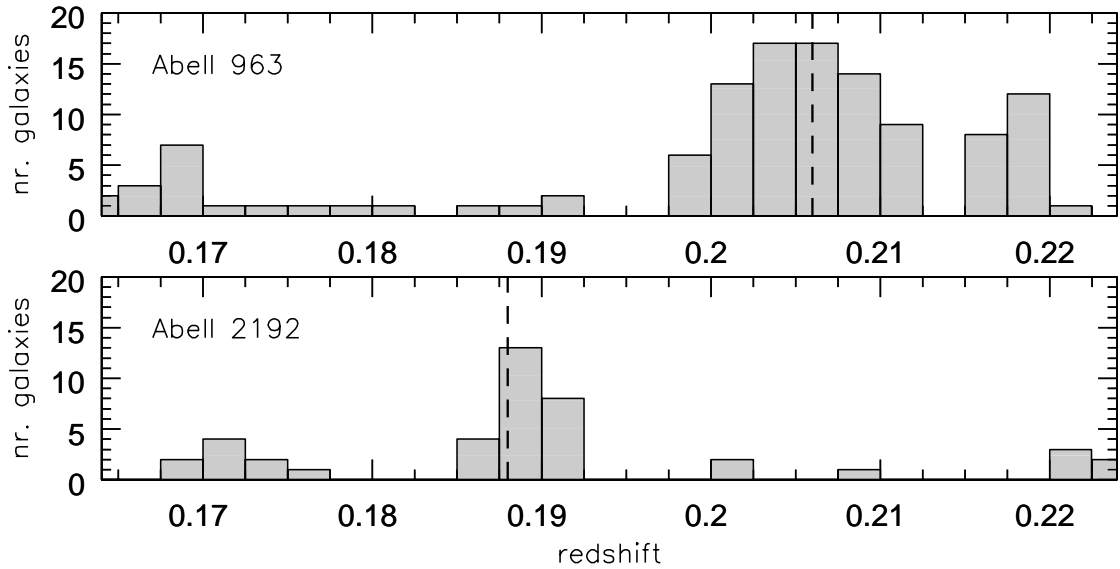}
\caption{The redshift distributions of the HI detected galaxies in
both fields. The vertical dashed lines indicate the redshifts of the
clusters based on optical spectroscopy of selected cluster
members. Note the overdensties in the field of Abell 963 at redshifts
of 0.167, 0.205 and 0.218, and in the field of Abell 2192 at redshifts
of 0.171, 0.189 and 0.221. This corresponds roughly to the large scale
structure within the surveyed volumes as shown in Figure 2.}
\label{fig6}
\end{figure}

Figure 4 shows random examples of twenty position-velocity diagrams
extracted from the data cubes at the lowest spectral resolution. These
slices were extracted at the positions where the HI emission peaks
while the angles at which these slices were taken do correspond to the
position angle of the identified optical counterpart. Despite the
large size of the synthesised beam, many sources are spatially
resolved, indicating extended HI disks. Note that not all sources
fullfil the detection criteria at this low spectral resolution. For
instance, the faintest sources in the upper and bottom rows exceed the
4$\times$3$\sigma$ threshold in the data cubes with the highest
spectral reslution while they have a UV-bright optical counterpart
within the FWHM of the synthesized beam.

Figure 5 presents in grayscales the total HI maps of all detections in
both surveyed volumes. In Abell 963, the projected density of HI
detections clearly peaks towards the center of the primary beam
although a few gas-rich galaxies are detected beyond the FWQM. In
Abell 2192, the distribution of HI detections in the field-of-view is
clearly asymmetric with most detections on the west side of the
cluster. This is obviously a manifestation of the large scale
structure within the surveyed volume and this seems to be corroborated
by a similar asymmetric distribution of galaxies with optical
redshifts from the SDSS. It should also be noted that not a single HI
detection was found near the center of Abell 963. Several HI aborption
systems against bright continuum sources have also been found but are
not identified in this figure.

\begin{figure}[t]
\begin{center}
\includegraphics[width=0.9\textwidth]{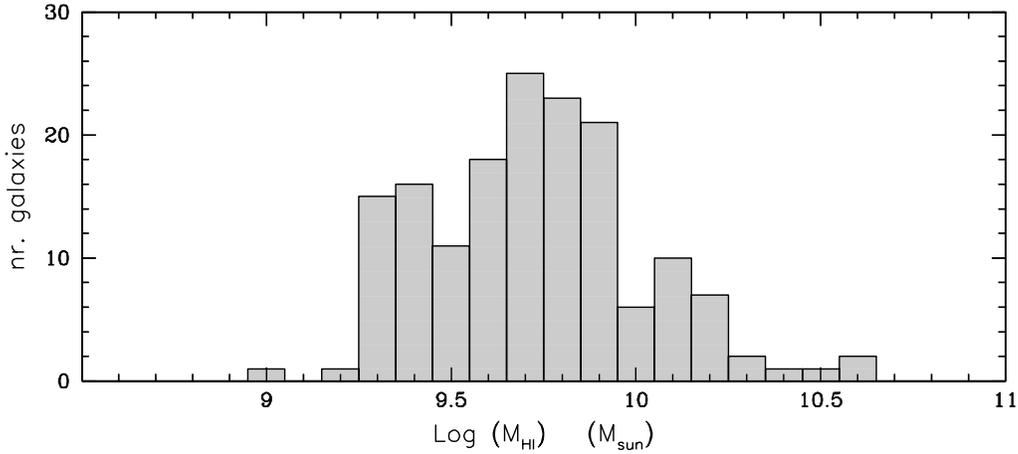}
\caption{The combined distribution of detected HI masses in both
volumes. The most gas-rich galaxies may be closely interacting galaxy
pairs that could not be resolved by the synthesized beam of the
WSRT.}
\label{fig7}
\end{center}
\end{figure}

Figure 6 depicts the redshift distributions of the HI detections in both
volumes. The redshifts of the clusters clearly stand out and the
overall distribution is in agreement with the distribution of optical
redshifts in these volumes as indicated in Figure 1. It is noteworthy
that the optical redshifts of the clusters, based on optical
spectroscopy of the brightest cluster members, do not coincide with
the peak of the redshift distributions of the HI detections. This may
indicate that the large scale structure in which these clusters are
embedded may be be asymmetric which is also evidenced by the total HI
map for the field centered on Abell 2192.

Figure 7 shows the distribution of the 160 measured HI masses. As
expected, the distribution peaks around M$_{\rm HI}^*$ with many
detections down to the HI mass limit of $\sim$2$\times$10$^9$
M$_\odot$. The distribution also has a remarkable tail toward the
highest HI masses. Some of these detections are associated with
closely interacting galaxies or compact galaxy groups that could not
be resolved by the synthesized beam of the WSRT. Inspection of the
optical images and the asymmetries of the global profiles will help to
resolve this issue. To construct the HI Mass Function and derive
$\Omega_{\rm HI}$ from these data requires thorough completeness
corrections and an assessment of the cosmic variance to estimate the
systematic uncertainties.

Figure 8 illustrates the relative gas content of the detected galaxies
as a function of their absolute R-band magnitude. At zero redshift, it
is well-known that the relative gas content of galaxies increases with
decreasing luminosity as illustrated by the open symbols. A similar
trend is observed at z$\approx$0.2 and at face value, this trend seems
to be even steeper with the fainter galaxies being even more
gas-rich. This, however, could be the result of the detection limit of
the observations as indicated by the slanted dashed line. To confirm
the suggestion that low-mass galaxies are more gas-rich at higher
redshifts, requires a careful investigation of the completeness near
the detection limit. This will be investigated in the near future.

Figure 9 presents a combined colour-magnitude diagram of all galaxies
with optical and HI redshifts in both volumes. Magnitudes and colours
from the INT observations are k-corrected and corrected for Galactic
exinction. No correction for internal extinction has been applied. The
HI-detected galaxies are predominantly located in the 'blue cloud'
while a few galaxies on the 'red sequence' also appear gas-rich. The
two brightest HI-detected galaxies are blue but at the same time as
bright as the most luminous galaxies on the red sequence. These are
possibly merging and star bursting galaxies but to address this issue
in any detail requires a study of the UV, optical and IR imaging data.

\begin{figure}[t]
\begin{center}
\includegraphics[width=0.9\textwidth]{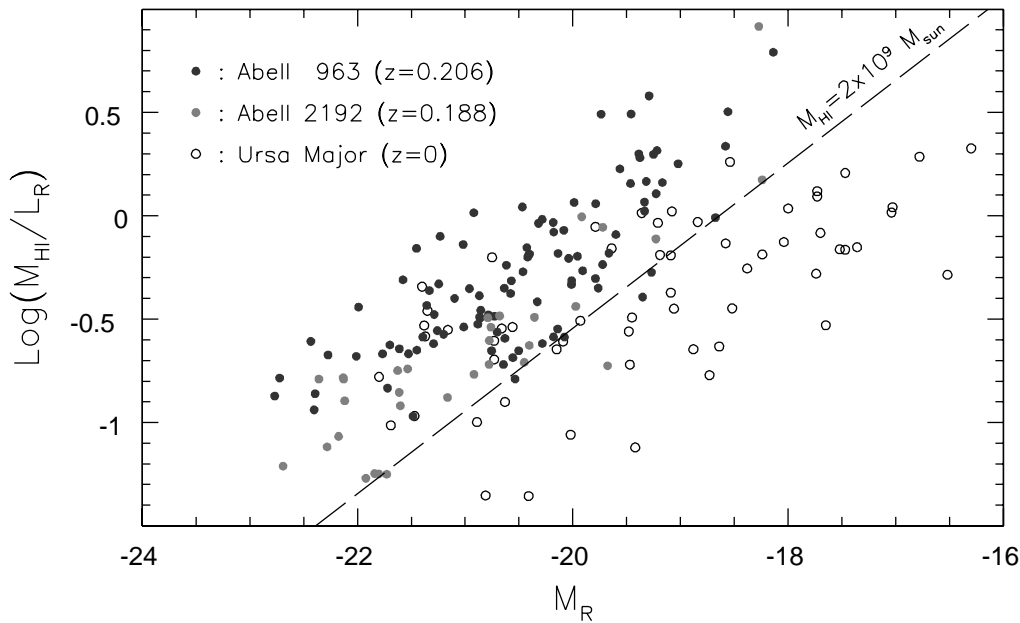}
\caption{The HI gas mass per solar luminosity of the HI detected
galaxies versus absolute magnitude, compared with a well-studied local
sample at z=0 (Verheijen \& Sancisi 2001). The slanted dashed line
indicates the nominal detection limit at the redshifts of the
clusters.}
\label{fig8}
\end{center}
\end{figure}

\begin{figure}[t]
\begin{center}
\includegraphics[width=0.9\textwidth]{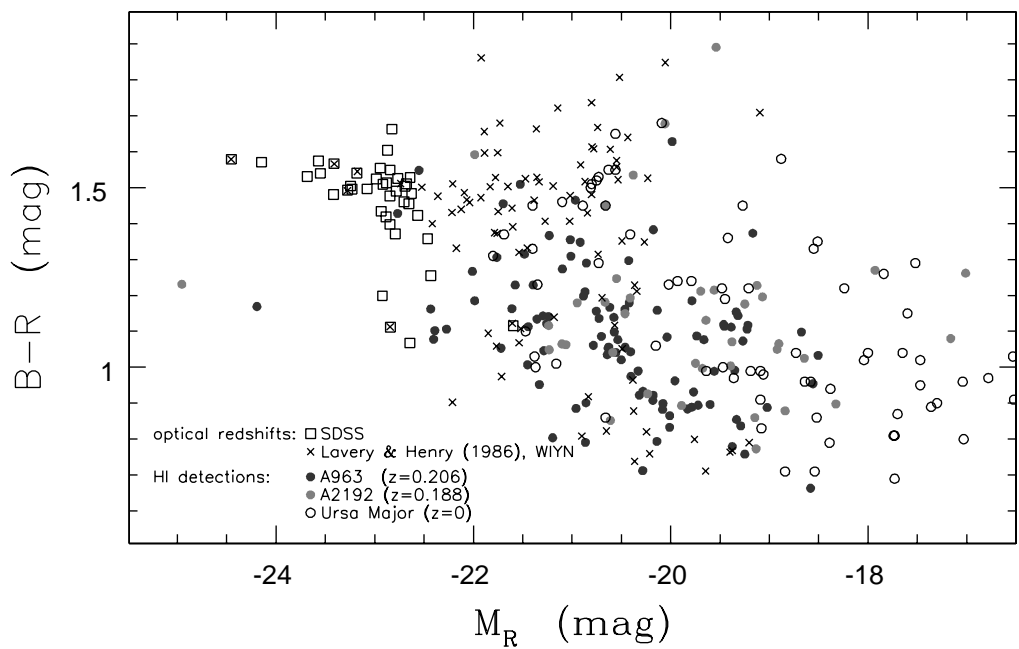}
\caption{K-corrected colour-magnitude diagram of all galaxies with
known redshifts (optical and HI) in both volumes. Galaxies targeted by
the SDSS are mainly located at the bright tip of the 'red
sequence'. HI detected galaxies are largely located within the 'blue
cloud' but some HI detections have very red optical counterparts.}
\label{fig9}
\end{center}
\end{figure}

\section{Conclusions}

We have demonstrated that observations of the HI 21cm emission line
from objects at z$\approx$0.2 are entirely feasible with present day
facilities. We have detected and measured HI emission from 160
galaxies in the redshift range 0.164$-$0.224 and revealed the cosmic
large scale structure outlined by these 160 galaxies. The HI detected
galaxies constitute a different population than the galaxies targeted
by the SDSS for optical spectroscopy at these distances. The largest
HI sources are spatially resolved by the WSRT while none of the blue
galaxies that are responsible for the B-O effect in Abell 963 have
been detected. There is a hint that the lower-luminosity galaxies at
z$\approx$0.2 are more gas-rich than their counterparts at low
redshift.

These WSRT observations have demonstrated that long integration times
with aperture synthesis arrays can reach the expected thermal noise in
the continuum subtracted channel maps.  Considering the surveyed
volume and the HI Mass Function in the Local Universe, the number of
160 HI detections in the WSRT survey is similar to the expected number
of 200 detections, modulo cosmic variance. This provides confidence in
the predicted number of HI detections in future ultra-deep surveys of
HI at intermediate redshifts with new facilities like MeerKAT,
APERTIF, ASKAP, EVLA, and the Square Kilometre Array.

\end{document}